\documentclass[conference]{IEEEtran}

\usepackage[short]{optidef}

\usepackage{pgfplots}
\usepackage{tikz}
\usetikzlibrary{calc}
 
\newcommand{\gettikzxy}[3]{%
  \tikz@scan@one@point\pgfutil@firstofone#1\relax
  \edef#2{\the\pgf@x}%
  \edef#3{\the\pgf@y}%
}

\usepackage{setspace}
\usepackage{lettrine}

\usepackage{graphics,
	psfrag,
	epsfig,
	amsthm,
	cite,
	amssymb,
	url,
	dsfont,
	subfigure,
	balance,
	enumerate,
	color,
	setspace
}

\usepackage{multirow}
\usepackage{algorithm} 
\usepackage{algorithmicx}
\usepackage{algpseudocode}

\usepackage{amsmath}

\usepackage{epstopdf}

\newcommand{\ignore}[1]{}

\usepackage{comment}
\usepackage{derivative}

\newcommand{\eref}[1]{(\ref{#1})}
\newcommand{\sref}[1]{Section~\ref{#1}}

\newcommand{\cref}[1]{Constraint~\ref{#1}}

\IEEEoverridecommandlockouts
\usetikzlibrary{spy,backgrounds}
\usetikzlibrary{shapes, arrows, positioning}

\newcommand{\ac}[1]{\textcolor{black}{#1}}

\usepackage{mathtools}
\usepackage{color}
\usepackage[utf8]{inputenc}

\begin{document}
\IEEEoverridecommandlockouts



\title{Uni-polarized RIS Beamforming for Improving Connectivity  of Multi-RIS-Assisted D2D Networks}

\author{\IEEEauthorblockN{Mohammed Saif, Mohammad Javad-Kalbasi, and Shahrokh Valaee}
\IEEEauthorblockA{Department of Electrical and Computer Engineering, University of Toronto, Toronto, Canada\\ 
Email: mohammed.saif@utoronto.ca, mohammad.javadkalbasi@mail.utoronto.ca, valaee@ece.utoronto.ca
\vspace{-0.35cm}
\thanks{This work was supported in part by funding from the Innovation for Defence Excellence and Security (IDEaS) program from the Department of National Defence (DND).}
}}
	
	

 

\maketitle

\begin{abstract}
This paper introduces a novel method to enhance the connectivity of multi-reconfigurable intelligent surface-assisted device-to-device networks, referred to as multi-RIS-assisted D2D networks, through a unique phase shift determination. The proposed method aims to optimize the power-domain array factor (PDAF), targeting specific azimuth angles of reliable user equipments (UEs) and  enhancing network connectivity. We formulate an optimization problem that jointly optimizes RIS beamforming design, RIS-aided link selection, and RIS positioning. This problem is a mixed-integer non-binary programming. The optimization problem is divided into two sub-problems, which are solved individually and iteratively. The first sub-problem of RIS-aided link selection is solved using an efficient perturbation method while  
developing genetic algorithm (GA) to obtain RIS beamforming design. The GA optimizes the RIS phase shift to generate multiple RIS-aided narrowbeams that exhibit significant PDAF towards   azimuth angles of interest while minimizing PDAF towards undesired azimuth angles. 
The second sub-problem of RIS positioning is addressed using the Adam optimizer. Numerical simulations verify the superiority of the
proposed scheme in improving   network connectivity compared to other schemes, including those utilizing distributed  small RISs, each generating one RIS-aided link.

\end{abstract}


\section{Introduction}

Reconfigurable intelligent surfaces (RISs) have emerged as a pivotal technology, enhancing various metrics of wireless networks, such as localization \cite{M}, energy efficiency \cite{Javad_globecom2023}, coverage \cite{RIS_Mohanad}, and network connectivity \cite{saifglobecom_E}. With their intelligent reflecting capabilities, RISs can be integrated with device-to-device (D2D) communications to connect  blocked UEs, thereby enhancing connectivity. 
To significantly enhance connectivity of D2D networks through RISs, it is crucial to optimize RIS beamforming judiciously, creating multiple cascaded links  (i.e., RIS-aided links) in the network. Specifically, RISs can improve network connectivity  by supporting direct links among UEs and connecting blocked UEs when direct links are unavailable, thus addressing the issue of zero connectivity. This issue arises when the network has more than one component \cite{saifglobecom_E}.

Network densification has been investigated to improve network connectivity through  deploying  many unmanned aerial vehicles (UAVs) \cite{8292633}, relays \cite{4786516},   and sensors \cite{4657335}. Deploying a large number of nodes in densely populated urban areas can be challenging due to site constraints and limited space and energy. Hence, RIS-assisted solution can overcome these limitations while maximizing the benefits of deploying more nodes without significantly impacting complexity or cost. Utilizing RISs to improve network connectivity is still not fully explored in the literature. In \cite{saifglobecom_E}, the authors  use RISs to improve connectivity of UAV networks  using matrix perturbation, each RIS generates a single narrowbeam RIS-aided link. Such study demonstrates good performance compared to RIS-free networks, which serves as a baseline scheme in this paper. Additionally, \cite{aydin} uses the RIS to boost the strength of the signals for resilient wireless networks. 
 
To advance the RIS beamforming optimization, this paper introduces a novel method to enhance the connectivity of multi-RIS-assisted D2D networks through a unique phase shift determination, such that it generates multiple narrowbeam RIS-aided links towards desired azimuth angles.  We formulate an optimization problem that jointly optimizes RIS beamforming design, RIS-aided link selection, and RIS positioning. The optimization problem is divided into two sub-problems and solved iteratively. The first sub-problem of RIS-aided link selection is solved using an efficient perturbation method while  developing genetic algorithm (GA) to obtain RIS beamforming design. The GA designs the RIS phase shift to generate multiple RIS-aided links that exhibit significant power-domain array factor (PDAF) towards   reliable UEs  while minimizing PDAF towards unreliable UEs. The second sub-problem of RIS positioning is addressed using the Adam optimizer. Numerical simulations verify the superiority of the
proposed scheme in improving   network connectivity compared to other  scenarios, including utilizing distributed  small RISs.

\section{System Model}\label{S}

\subsection{Network Model}\label{NM}
Consider a 2D multi-RIS-assisted D2D model that consists of  multiple UEs, denoted by the set $\mathcal U=\{1, 2, \ldots, U\}$  and multiple RISs, denoted by the set $\mathcal M=\{1, 2, \ldots, M\}$. In a dense urban scenario, where direct links between transmitting UEs and receiving UEs are blocked, communication can only occur through the RISs. RISs can significantly aid in establishing reliable communication with the blocked receiving UEs. This work considers a general scenario where direct links between the UEs might be unavailable. Thus, RISs can enhance  network connectivity by supporting direct links of UEs and connecting unconnected UEs. It is assumed that the UEs are equipped with a single antenna, while an RIS has $N$ reflecting elements with a horizontal uniform linear array (ULA) topology.  In an RIS, each meta-atom includes a reflector that not only reflects signals but can also independently adjust the phase of the incoming wireless signals. We denote $\mathbf{\Phi}^{m}=[\phi_{1}^{m},\ldots, \phi_{n}^{m}, \ldots, \phi_{N}^{m}]$ as the phase shift design vector of  RIS$_m$, where $\phi_{n}^{m}$ is the phase shift induced by the $n$-th RIS element to the incoming signal. Let  $\mathbf{\Psi}^{m}=[\psi_{1}^{m},\ldots,\psi_{n}^{m},\ldots,\psi_{N}^{m}]$ be a vector with $\psi_{n}^{m}=e^{j\phi_{n}^{m}}$.

 

This work considers that each RIS reflects the signal of a typical transmitting UE to multiple receiving UEs through \ac{beamforming}; however, \ac{an} RIS cannot be assigned to
more than one transmitting UE at the same time. This assumption is justified in \cite{9293155}. Thus, the set of the possible transmitting UEs, denoted by $\mathcal U_t$, is at most $\mathcal M$. The set of multiple receiving UEs that exploit RIS$_m$ is denoted as $\mathcal U_m$, which also represents the number of beamformers of RIS$_m$, i.e., $U_m=|\mathcal U_m|$. 
The RIS beamforming needs to be designed  
to concurrently boost the signal dedicated to the least critical receiving UEs (i.e., reliable UEs).  High critical UEs (i.e., unreliable UEs), which need to be avoided, are the most critical ones that cause severe connectivity degradation if they fail. The reliability of the UEs will be defined in \sref{NC}.

\subsection{Channel Model and SINR Formulation} \label{CM}
All channels   are considered to be quasi-static and presumed to be perfectly-known. Let $\mathbf h^{m}_u \in \mathbb C^N$ and $\mathbf g^{m}_r \in \mathbb C^N$ represent the transmitting UE-RIS and RIS-receiving UE channels, respectively. For simplicity, we consider a line-of-sight (LoS) channel between the transmitting UEs and the RISs and between the RISs and the receiving UEs. The LoS channels between  UE$_u$ and RIS$_m$ and between RIS$_m$ and  UE$_r$ can be expressed, respectively,  as $\mathbf h^{m}_u=~\sqrt{\beta^{m}_u G_0(\theta^{m}_u)} \mathbf a(\theta^{m}_u)$ and $\mathbf g^{m}_r= \sqrt{\beta^{m}_r G_0(\theta^{m}_r)} \mathbf a {(\theta^{m}_r)}$, where $\beta^{m}_u$ and $\beta^{m}_r$ denote the corresponding path-losses of $\text{UE}_u \rightarrow \text{RIS}_m$ and $\text{RIS}_m \rightarrow \text{UE}_r$ links, respectively,  $\theta_{u}^{m}$ represents the \ac{angle-of-arrival (AoA)} from  UE$_u$ to RIS$_m$, which is assumed to be known and maintained constant, $\theta^{m}_r$ is the \ac{angle-of-departure (AoD)} from RIS$_m$ to  UE$_r$, which is the azimuth angle of UE$_r$, $G_0(\cdot)$ is the radiation power pattern of a single RIS element, and $\mathbf a(\cdot)$ is the RIS array response vector, which can be expressed as \cite{arrayresp}
 $\mathbf{a}(x)=[1,e^{-j\frac{2\pi \Delta}{\lambda} \sin{x}}, \ldots,  e^{-j\frac{(N-1)2\pi \Delta}{\lambda} \sin{x}}]^T$, where $\Delta$ is the spacing between the adjacent RIS elements (i.e., inter-element spacing) and $\lambda$ is the wavelength of the transmitted signal.   For D2D channels, let $h^\text{U}_{u,r}$ and $\beta^\text{U}_{u,r}$ denote the small-scale fading coefficient and path-loss for the $\text{UE}_u \rightarrow \text{U}_r$ channel, respectively.

For a reference $\text{UE}_u \xrightarrow{\text{RIS}_m} \text{UE}_{r}$, the signal received at UE$_r$ can be written as
\begin{align}\label{RS}  
y^{m}_r=  y^{m}_{u,r} + \sum_{\substack{u'\in \mathcal U_t, u' \neq u}} y^{m'}_{u',r} + \zeta_r,
\end{align}
where the first term is the  signal received from UE$_u$, the second term is the signal received from the other transmitting UEs over the other RISs (i.e., $u' \neq u$ and $m' \neq m$), and the third term $\zeta_r$ is the additive white Gaussian noise (AWGN)  at UE$_r$ with $w_r \sim  \mathcal {CN}(0, \sigma^2_{\zeta_r})$, where $\sigma^2_{\zeta_r}$ is the variance. The first   term  of \eref{RS} can be expressed as follows 
\begin{align} \label{RS1}
y^{m}_{u,r}=\bigg(\underbrace{\sqrt{\beta^\text{U}_{u,r}}h^\text{U}_{u,r}}_\textbf{Direct Link of UE$_u$}+ \underbrace{(\mathbf g^{m}_r)^T \text{diag}\left(\mathbf{\Psi}^{m}\right) \mathbf h^{m}_u}_\textbf{Signal From RIS$_m$} \bigg)\sqrt{p}x_u,
\end{align}
and  $y^{m'}_{u',r}$ is same as $y^{m}_{u,r}$ but for $u'$ and $m'$, where $\text{diag}\left(\mathbf{\Psi}^{m}\right)$ is a diagonal matrix with diagonal elements $\mathbf{\Psi}^{m}$ and $x_u$ is the transmitted signal of UE$_u$. The signal received at UE$_r$ in \eref{RS} has the following components:
\begin{itemize}

\item Signal received from the direct link, $\sqrt{\beta^\text{U}_{u,r}}h^\text{U}_{u,r}$

\item Signal received from RIS$_m$ to UE$_r$, $(\mathbf g^{m}_r)^T \text{diag}\left(\mathbf{\Psi}^{m}\right) \mathbf h^{m}_u$

\item Signal received from the direct link of other transmitting UEs$_{u'}$, $\sqrt{\beta^\text{U}_{u',r}}h^\text{U}_{u',r}$

\item Lastly, signal received from other transmitting UEs$_{u'}$ over other RIS$_{m'}$ ($m' \neq m$) to UE$_r$.  
\end{itemize}
We can alternatively rewrite \eref{RS} as \eref{FL_1} given at the top of the next page, where the notation \(\mathbf{a}(\theta^{m}_{u}) \odot \mathbf{a}(\theta^{m}_{r})\) represents the element-wise product between the vectors \(\mathbf{a}(\theta^{m}_{u})\) and \(\mathbf{a}(\theta^{m}_{r})\) and $p$ is the transmit power of UE$_u$. 
\begin{table*}
	\begin{align}
	\label{FL_1} \nonumber 
	y^{m}_r=&\bigg(\sqrt{\beta^\text{U}_{u,r}}h^\text{U}_{u,r}+\sqrt{G_0(\theta^{m}_{u}) G_0(\theta^{m}_{r})\beta^{m}_{u} \beta^{m}_{r}} \mathbf{\Psi}^{m}   (\mathbf{a}(\theta^{m}_{u})\odot\mathbf{a}(\theta^{m}_{r})) \bigg)\sqrt{p}x_u + \sum_{\substack{u'\in \mathcal U_t, u' \neq u}}\bigg(  \sqrt{\beta^\text{U}_{u',r}}h^\text{U}_{u',r}\\&   +\sqrt{G_0(\theta^{m'}_{u'}) G_0(\theta^{m'}_{r})\beta^{m'}_{u'} \beta^{m'}_{r}} \mathbf{\Psi}^{m'}   (\mathbf{a}(\theta^{m'}_{u'})\odot\mathbf{a}(\theta^{m'}_{r})) \bigg)\sqrt{p}x_{u'} + \zeta_r, 
	\end{align}  
\end{table*}
\begin{table*}
	\begin{align}\label{SINR} 
\gamma^{m}_{u,r}(\boldsymbol{\alpha}^{m})=\frac{p\bigg[ X^{(d)}_u+X^{(c)}_u \mathbf{\Psi}^{m}   (\mathbf{a}(\theta^{m}_{u})\odot\mathbf{a}(\theta^{m}_{r}))\bigg]^2}{p\sum_{\substack{u'\in \mathcal U_t\\ u' \neq u}}\bigg[ X^{(d)}_{u'}+X^{(c)}_{u'}\mathbf{\Psi}^{m'}  (\mathbf{a}(\theta^{m'}_{u'})\odot\mathbf{a}(\theta^{m'}_{r}))\bigg]^2+\sigma^2_{\zeta_r}},
\end{align}
	\hrulefill
	\vspace*{-0.5cm}
\end{table*}

Let $X^{(d)}_u=\sqrt{\beta^\text{U}_{u,r}}h^\text{U}_{u,r}$, $X^{(d)}_{u'}=\sqrt{\beta^\text{U}_{u',r}}h^\text{U}_{u',r}$, $X^{(c)}_u=\sqrt{\beta^{m}_{u} \beta^{m}_{r} G_0(\theta^{m}_{u}) G_0(\theta^{m}_{r})}$, and $X^{(c)}_{u'}=\sqrt{\beta^{m'}_{u'} \beta^{m'}_{r} G_0(\theta^{m'}_{u'}) G_0(\theta^{m'}_{r})}$,  then
the signal-to-interference plus noise ratio (SINR) at UE$_r$  can be written as  \eref{SINR} given at the top of the next page, where $\boldsymbol{\alpha}^m=[\alpha^{m}_x, \alpha^{m}_y]$ is the Cartesian
coordinates of RIS$_m$.

Note that in \eref{SINR}, the received signal power from all the other RISs (i.e., RIS$_{m'}, m'\neq m, \forall m' \in \mathcal M$)  can be neglected compared to the received signal power from the aligned RIS$_m$ and the direct links. Thus, to ease the analysis for UE-RIS-UE optimization in  \sref{PS}, we make an approximation for the SINR expression in \eref{SINR} by ignoring the   term $X^{(c)}_{u'}\mathbf{\Psi}^{(m')}  (\mathbf{a}(\theta^{m'}_{u'})\odot\mathbf{a}(\theta^{m'}_{r}))$.
Accordingly, the SINR at UE$_r$ can be mathematically approximated as
\begin{equation}\label{app}
\gamma^{m}_{u,r}(\boldsymbol{\alpha}^{m})=\frac{p\bigg[ X^{(d)}_u+X^{(c)}_{u}\mathbf{\Psi}^{m}  (\mathbf{a}(\theta^{m}_{u})\odot\mathbf{a}(\theta^{m}_{r}))\bigg]^2}{p\sum_{\substack{u'\in \mathcal U_t\\ u' \neq u}}\bigg[ X^{(d)}_{u'}\bigg]^2+\sigma^2_{\zeta_r}},
\end{equation}
The accuracy of this approximation\ac{, in regards to the exact SINR expression in \eref{SINR},} is verified via numerical simulations in \sref{NR}.

\section{Problem Modeling}\label{PF}

\subsection{Network Connectivity and Node Reliability} \label{NC}
We model the connectivity of the D2D system (without RISs) using the graph network $\mathcal G(\mathcal V, \mathcal E)$, where $\mathcal V$ represents the set of vertices associated with the UEs and $\mathcal E$ represents edges associated with the D2D links. To model the weight of the D2D link $\text{UE}_u \rightarrow \text{UE}_r$, we consider the signal-to-noise ratio (SNR), which is defined as  $\gamma^\text{U}_{u,r}=\frac{p|\sqrt{\beta^\text{U}_{u,r}}h^\text{U}_{u,r}|^2}{N_0}$, where $N_0$ is AWGN variance. Let $\gamma^\text{U}_{0}$ \ac{be} the minimum SNR threshold for the  D2D links, then edge $e_{l}$ connects two vertices $(v, v') \in \mathcal V$, if  $\gamma^\text{U}_{v,v'} \geq \gamma_0^\text{U}$, \ac{and 
otherwise, they are not connected.}  

The weight vector $\mathbf w \in {[\mathbb R^+]}^E$ of the D2D links is defined as $\mathbf w= [ w_1, w_2, \ldots, w_E]$, which is given element-wise as
$w_l=w_{v,v'}=\gamma^\text{U}_{v,v'}$. For $e_{l}$, let $\mathbf a_l$ be a vector, where the $v$-th and the $v'$-th elements in $\mathbf a_l$ are given by $a_{v,l}=1$ and $a_{v',l}=-1$, respectively, and zero otherwise. Let $\mathbf A$ be the incidence matrix of a graph $\mathcal G$ with the $l$-th column given by $\mathbf a_l$.  The Laplacian matrix $\mathbf L$ is a $V \times V$ matrix, defined as \cite{4657335}
\begin{equation} \label{lap}
\mathbf L= \mathbf A  ~diag(\mathbf w) ~\mathbf A^T=\sum^{E}_{l=1} w_l \mathbf a_l \mathbf a^T_l,
\end{equation}
where the entries of $\mathbf L$ are given by
\begin{equation}
L(v,v') = \begin{cases}
\sum_{\Tilde{v} \neq v} w_{v, \Tilde{v}} &\text{if} ~v=v',\\
-w_{v,v'} &\text{if}~ (v, v') \in \mathcal E, \\
0 & \text{otherwise}.
\end{cases}
\end{equation}

Similar to \cite{new, 4657335,   4786516, 8292633}, we choose  the \textit{algebraic connectivity}, also called the Fiedler value, denoted as $\lambda_2(\mathbf L)$, to measure the connectivity of the considered network. With RIS-aided link deployment, a new graph $\mathcal G'(\mathcal V, \mathcal E')$ has a larger set of edges denoted by $\mathcal E'$ with $\mathcal E'=\mathcal E \cup \mathcal E_{new}$, where $\mathcal E_{new}$ is the new  
$\text{UE}_u \xrightarrow{\text{RIS}_m} \text{UE}_{r}$ edges, $\forall u\in \mathcal U_t, r \in \mathcal U_m, m\in \mathcal M$. The  gain of RIS deployment on network connectivity can be assessed by computing $\lambda_2 (\mathbf L') \geq \lambda_2 (\mathbf L)$, where $\mathbf L'$ is the resulting Laplacian matrix of the new graph $\mathcal G'(\mathcal V, \mathcal E')$.

Let $\mathcal G_{-v}$ be the sub-graph obtained after removing vertex $v \in \mathcal V$ along with all its adjacent edges to other vertices in $\mathcal G$, i.e., $\mathcal G_{-v} \subseteq \mathcal G$.   The connectivity of $\mathcal G_{-v}$  is defined as $\lambda_2(\mathcal G_{-v})$. A node that, when removed, significantly reduces the network connectivity  is declared to be highly critical and thus not reliable.   Therefore,  we measure the reliability of the nodes based on their criticalities, which reflects the  severity of the impact on the connectivity of  $\mathcal G_{-v}$, which is defined  as 
\begin{equation}\label{eq1}
\mathcal R_v=\lambda_2(\mathcal G_{-v}).
\end{equation}
Equation \eref{eq1} implies that \ac{highly} critical nodes are not reliable, i.e., if $\mathcal R_v > \mathcal R_{v'}$, node $v$ has higher reliability than node $v'$. 

\subsection{Problem Formulation}
\begin{figure}[t!]  
\begin{center}
\includegraphics[width=0.75 \linewidth, draft=false]{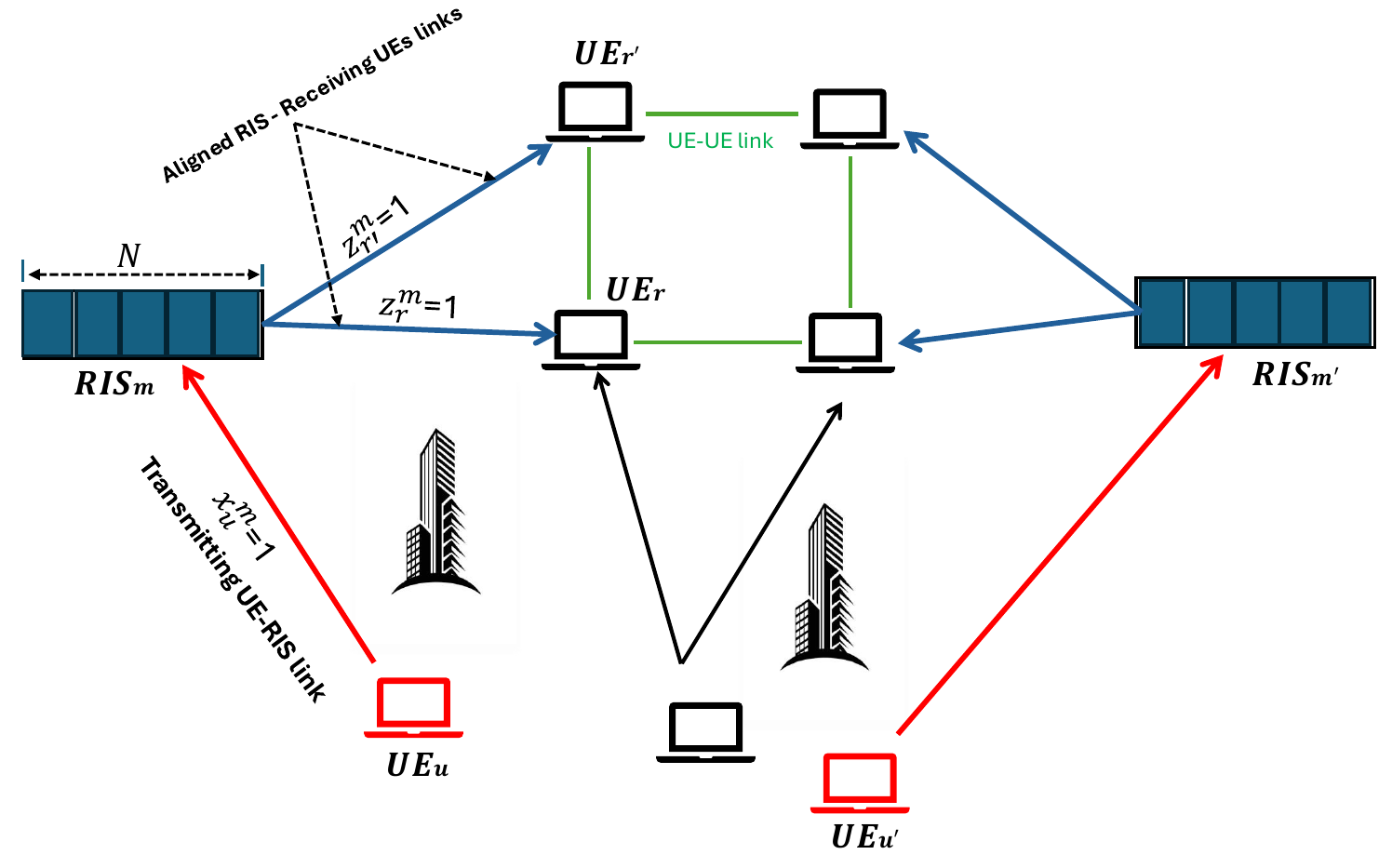}
\caption{\ac{Network configuration of a multi-RIS assisted D2D model, clarifying the connections of the transmitting UE$_u$ and RIS$_m$ for $U_m=2$.}}
   \label{fig11}
\end{center}
 
\end{figure}

Let $z^{m}_{r}$ be a binary variable  that is $1$  if \ac{the receiving} UE$_r$ is connected to RIS$_m$, and $0$ otherwise. The binary RIS-receiving UE assignment matrix is symbolized as $\mathbf Z$. Likewise, let $x^{m}_u$ be binary variable that is $1$ if  \ac{the transmitting} $\text{UE}_u$ is connected to RIS$_m$, and $0$ otherwise, and the binary transmitting UE-RIS assignment matrix is symbolized as $\mathbf X$. \ac{The network configuration that shows these binary variables is presented in Fig.~\ref{fig11}.}

Let  $\gamma^\text{RIS}_{0}$ be the minimum SINR threshold of  $\text{UE}_u \rightarrow \text{UE}_{r}$  via RIS$_m$.   The \ac{proposed} optimization problem is formulated as  
\begin{subequations} \nonumber 
\label{eq10}
\begin{align}
&\mathcal P_0: ~\max_{\boldsymbol{\alpha}, \mathbf X, \mathbf Z, \mathbf{\Phi}} ~~~~ \lambda_2(\mathbf L' (\boldsymbol{\alpha}, \mathbf X, \mathbf Z, \mathbf{\Phi}))
\label{eq10a}\\
 & {\rm s.~t.\ } \text{C$_1^0$:} ~ \sum_{m \in \mathcal M} z^{m}_{r} \leq 1, ~~~~~~~~~~~   \forall r \in \{1, 2, \dots,  U\},\\
  &~~~~~~\text{C$_2^0$:} ~\sum_{m \in \mathcal M} \sum_{u \in \mathcal U_t} x^{m}_u \leq M,\\
  &~~~~~~\text{C$_3^0$:} ~ x^{m}_u\sum_{r \in \mathcal U}  z^{m}_{r}  \leq U_m, ~~~~~~~ \forall m \in \mathcal M, \forall u \in \mathcal U_t,\\
 &~~~~~~\text{C$_4^0$:} ~ \gamma^{m}_{u,r}(\boldsymbol{\alpha}^{m}, \mathbf X, \mathbf Z, \mathbf \Phi^{m})\geq \mathcal R_r\gamma^\text{RIS}_{0},  ~\forall (u,m,r), \\
&~~~~~~\text{C$_5^0$:}~ \phi^{m}_n \in [0, 2\pi),~~~   \forall m \in \mathcal M, n=\{1, \ldots, N\},\\
& ~~~~~~~~   z^{m}_{r}, x^{m}_u \in \{0,1\}, ~~~~~\forall u \in \mathcal U_t, m \in \mathcal M, r \in \mathcal U,
\end{align}
\end{subequations} 
where  $\boldsymbol{\alpha}=[\boldsymbol{\alpha}^{1}, \ldots, \boldsymbol{\alpha}^{M}]$ and $\boldsymbol{\Phi}=[\boldsymbol{\Phi}^{1}, \ldots, \boldsymbol{\Phi}^{M}]$. In $\mathcal P_0$, \text{C$_1^0$} implies that at most one reflected link should be created for UE$_r$  via the RISs. \text{C$_2^0$} shows that at most $M$ links are created between the transmitting UEs and the RISs. \text{C$_3^0$} enforces that each RIS is involved in reflecting the signal of a \ac{single} transmitting UE to at most $U_m$ receiving UEs. This also means  that, at most, $U_mM$ RIS-aided links can be created in the network.  \text{C$_4^0$}  
constitutes the  \ac{quality-of-service} (QoS) constraint on receiving UE$_r$ based on its reliability. Specifically,  the reliability metric $\mathcal R_r$ controls the QoS limit set for receiving UE$_r$, $\forall r$.  Finally, \text{C$_5^0$} is for the RIS phase shift optimization.

\section{Proposed Solution}\label{PS}
\begin{figure}[t!]  
\begin{center}
\includegraphics[width=0.55 \linewidth, draft=false]{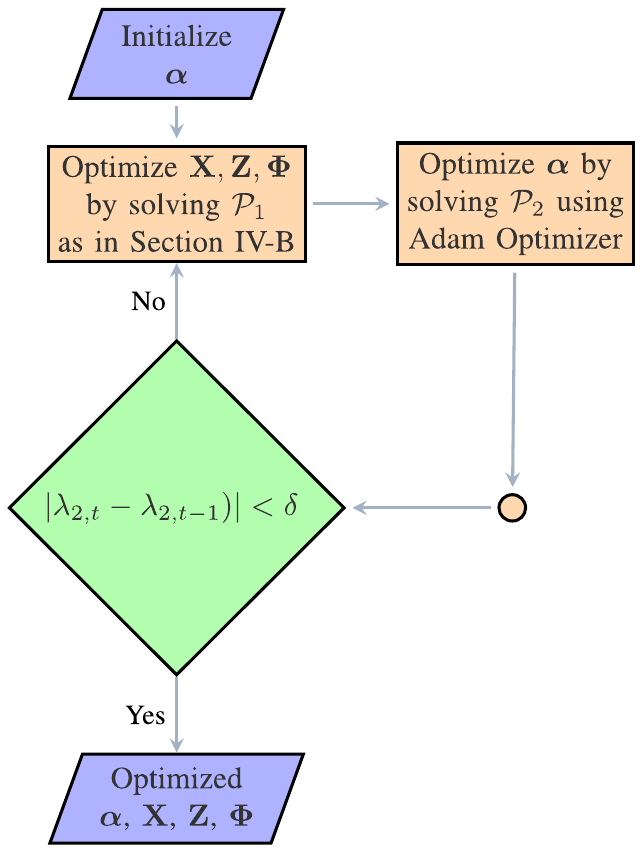}
\caption{The flowchart for the iterative solution.}
   \label{Fig:block_diagram_bcd}
\end{center}
 
\end{figure}

The derived optimization problem $\mathcal P_0$  is a mixed-integer non-binary programming. 
This section presents the proposed iterative solution  to tackle $\mathcal P_0$ for $\mathbf X$, $\mathbf Z$, $\mathbf \Phi$, and $\boldsymbol{\alpha}$ as presented in Fig.~\ref{Fig:block_diagram_bcd}. 
 
\subsection{Narrowband Beamforming Design for $\mathbf{\Phi}$}\label{CGAA}

To generate a narrowband beamforming for $\text{UE}_u \xrightarrow{\text{RIS}_m} \text{UE}_{r}$ link, we consider the following  PDAF of RIS$_m$  \cite{Uni}  
\begin{align}\label{PDAF} \nonumber
A^{m}(\mathbf{\Phi}^{m},\theta^{m}_{r})&=   {\left| \mathbf{\Psi}^{m}(\mathbf{a}(\theta^{m}_{u})\odot \mathbf{a}(\theta^{m}_{r}))\right | }^{2}
 \\ & = {\left |\sum\limits_{n=1}^{N} e^{j\phi_{n}^{m}}e^{-j\frac{2\pi \Delta (n-1)}{\lambda}\left(\sin{\theta^{m}_{u}}+\sin{\theta^{m}_{r}}\right)} \right |}^{2}.   
\end{align}
To design $\mathbf{\Phi}^{m}$ that generates $U_m$ narrowband beamforming to the desired azimuth angles of $U_m$ UEs from RIS$_m$ while maximizing the minimum PDAF, our goal is to address the following optimization problem
\begin{equation}\label{asliopt}
\max_{\mathbf{\Phi}^{m}} \min_{\theta \in \{\theta^{m}_{1}, \dots, \theta^{m}_{U_m}\}} A^{m}(\mathbf{\Phi}^{m},\theta).
\end{equation}

\begin{algorithm}[t]
 \begin{algorithmic}[1]
 \renewcommand{\algorithmicrequire}{\textbf{Input: Fitness function $\eta(\boldsymbol{\theta},\boldsymbol{P})$ and population size $2J$}}
 \State  Generate the initial population $\mathbb{P}^{m,0}= {\left[\mathbf{\Phi}_{1}^{m,0},\mathbf{\Phi}_{2}^{m,0},\ldots,\mathbf{\Phi}_{2J}^{m,0}\right]}^{T}$.
 \State  Let $\Gamma\left({\mathbf{\Phi}_{j}^{m,0}}\right)=\min\limits_{\theta \in \{\theta^{m}_{1}, \dots, \theta^{m}_{U_m}\} }A\left(\mathbf{\Phi}_{j}^{m,0},\theta \right)$ be the fitness value for $j=1,2,\ldots,2J$.
 \State  Define the selection random variable $Q^{m,0}$ on the set $\{1,2,\ldots,2L\}$ with probability mass function $Pr\{Q^{m,0}=k\}=\frac{\Gamma\left({\mathbf{\Phi}_{k}^{m,0}}\right)}{\sum\limits_{j=1}^{2J}\Gamma\left({\mathbf{\Phi}_{j}^{m,0}}\right)}$.
 \For {$c=1:C$}
 \State  Select \( 2J \) individuals from the set \( \mathbb{P}^{c-1} \) based on  the distribution that is defined by \( Q^{m, c-1} \).
 \State  Randomly select \( J \) pairs of individuals from \( \mathbb{P}^{m,c-1} \) for crossover. 
 \State  For mutation, add a small random perturbation to each element of the previously generated offsprings.
 \State  Obtain the updated population \( \mathbb{P}^{m,c} \) and the corresponding  \( Q^{m,c} \).
 \EndFor
 \end{algorithmic} 
 \caption{Proposed  GA for RIS$_m$}
 \label{CGA}
 \end{algorithm}

We solve this optimization problem by utilizing the \ac{genetic algorithm (GA)}, which is an optimization technique for finding effective solutions \cite{Lit13}. Inspired by natural selection and evolutionary processes, the GA evolves a pool of potential solutions over successive generations, progressively refining them. The main steps of the GA are as follows: 
\begin{enumerate}

\item \textbf{Initialization}: The GA randomly generates population for initial solutions, where the size of population is determined by the problem complexity. We represent each  candidate solution by a vector of continuous values.

\item \textbf{Selection}: The candidate solutions are then assessed using a fitness function that assesses their performance on the problem. In this paper, $\Gamma\left({\mathbf{\Phi} }\right)\triangleq \min_{\theta \in \{\theta^{m}_{1}, \dots, \theta^{m}_{U_m}\}} A^{m}(\mathbf{\Phi}^{m},\theta)$ is the fitness function. The best candidate solutions are nominated as parents for the next generation.

\item \textbf{Crossover}: By combining pairs of parent solutions, the GA produces offspring solutions via calculating the weighted sum of the parent vectors, resulting in new solution vectors. Crossover helps in exploring the search space more comprehensively to generate a variety of candidate solutions.

\item \textbf{Mutation}: The GA may \ac{get} stuck in   local optima, and to avoid this while diversify the search, the GA performs random alterations to the offspring solutions; adding a small random value to each element of the solution vector, ensuring  the phase shift constraint. 
\end{enumerate}

The GA iterates through the above steps until it satisfies a specified termination criterion. Algorithm \ref{CGA} provides detailed implementation of the GA.


\subsection{Iterative Solution}\label{PS}
When the RIS placement is fixed (i.e., $\boldsymbol{\alpha}^{m}=\boldsymbol{\alpha}^{m}_0$, $\forall m$), the sub-problem for optimizing $\mathbf X$ and $\mathbf Z$ while utilizing the designed GA for $\mathbf \Phi$ is given by
\begin{subequations} \nonumber 
 
\begin{align}
&\mathcal P_1: ~\max_{\mathbf X, \mathbf Z, \mathbf \Phi} ~~~~ \lambda_2(\mathbf L' (\mathbf X, \mathbf Z, \mathbf \Phi))
\label{eq10a}\\
 & {\rm s.~t.\ }~
 \text{C$_1^0$},~ \text{C$_2^0$},~ \text{C$_3^0$}, ~\text{C$_5^0$},\\
 &~~~~~~~\text{C$_4^0$:} ~ \gamma^{m}_{u,r}(\boldsymbol{\alpha}^{m}_0, \mathbf X, \mathbf Z, \mathbf \Phi^{m})\geq \mathcal R_r\gamma^\text{RIS}_{0}.
\end{align}
\end{subequations} 

For optimizing $\mathbf X$ and $\mathbf Z$, we propose an effective greedy perturbation  based on the values of the Fiedler vector  of the original network. This vector is denoted by $\mathbf v$, which is the eigenvector corresponding to the Fiedler value of 
$\mathbf L$ that provides valuable information about the connectivity of the graph.  Let $W_{ur}=(v_u-v_r)^2$, where  $v_u$ and $v_r$ are the corresponding values of UE$_u$ and UE$_r$ indices of the Fiedler vector $\mathbf v$ of $\lambda_2 (\mathbf L)$. Thus,  we propose to weight these differences of $W_{ur}$ by $w_rW_{ur}$,  where $w_r$ is the weight defined in \eref{eq1} that represents the reliability of UE$_r$. The weighted differences of $w_rW_{ur}$ indicate the connection strength between the UEs \cite{new}, meaning  large differences suggest that connecting the corresponding edge would significantly enhance connectivity while selecting the most reliable receiving UE. Conversely, small differences yield moderate connectivity improvement. By analyzing these differences, the perturbation method can prioritize which UEs   need to be connected via the RISs.

Each step of the proposed perturbation  chooses an edge $l$ that connects UE$_u$ and UE$_r$, which has the largest value of $w_{r}W_{ur}$ that maximizes $\lambda_2(\mathbf L')$. Beginning with $\mathcal G$ and $\mathbf L$, the steps of the proposed perturbation are given in Algorithm \ref{Algorithmper}.   

\begin{algorithm}[t!]
	\caption{The Perturbation Method for solving $\mathcal P_1$}
	\label{Algorithmper}
	\begin{algorithmic}[1]
		\State \textbf{Input:} $\boldsymbol{\alpha}$ and $\mathcal G$  
  \State Initially set $\mathbf L' \leftarrow \mathbf L$
   \State Calculate $w_r$ for all UEs based on \eref{eq1}, $\forall r$
        \For{$m=1, 2, \ldots, M$}
        \State Calculate the Fiedler vector $\mathbf v$ of the associated $\mathbf L'$ 

        \State Calculate $w_{r}W_{ur}, \forall (u, r) \in \mathcal U$
        \State From the remaining edges, add an
edge $l$ connecting UE$_u$ and UE$_r$ with largest $w_{r}W_{ur}$
         \State With the same UE$_u$, continue Step $7$ for $U_m-1$ links
         \State For the  $U_m$ edges, design $\mathbf \Phi^{m}$ using Algorithm \ref{CGA}
        \State Given the selected edges, update $\mathbf L'$
         \State Remove all the RIS-aided candidate links of the already selected nodes
         \EndFor         
	\State \textbf{Output:} $\mathbf X, \mathbf Z$
	\end{algorithmic}
\end{algorithm}

When $\mathbf X$, $\mathbf Z$, and $\mathbf \Phi$ are fixed, the sub-problem for optimizing $\boldsymbol{\alpha}$ is given by
\begin{subequations} \nonumber 
\begin{align}
&\mathcal P_2: ~\max_{\boldsymbol{\alpha}} ~~~~ \lambda_2(\mathbf L' (\boldsymbol{\alpha}))
\label{eq10a}\\
 & {\rm s.~t.\ }~
 \text{C$_4^0$:} ~\gamma^{m}_{u,r}(\boldsymbol{\alpha}^{m})\geq \mathcal R_r\gamma^\text{RIS}_{0}, ~~~~~~~~~~\forall (u,r,m).
\end{align}
\end{subequations} 

To optimize the   RIS positioning, we use the Adam optimizer, which is a gradient-based optimization algorithm that is used in machine learning due to its superior performance compared to other methods \cite{adam}. The detailed steps of the Adam optimizer  are given in Algorithm \ref{Algorithm4}.


 \begin{algorithm}[t!]
	\caption{Adam Optimizer for solving $\mathcal P_2$}
	\label{Algorithm4}
	\begin{algorithmic}[1]
		\State \textbf{Input:} $\mathbf X$, $\mathbf Z$, step size $\nu$, small constant  $\epsilon$,  exponential decay rates for the moment estimates $\beta_{1},\beta_{2} \in [0,1)$, and initial position of RISs $\boldsymbol{\alpha}_0$
        \State \textbf{Initialize:} $m_0 \leftarrow 0$, $v_0  \leftarrow  0$, and $i  \leftarrow  0$
        \For{$i=1:I$}

         \State $g_i \gets \nabla_{\boldsymbol{\alpha}}\lambda_2(\mathbf L'\left(\boldsymbol{\alpha}_{i-1}\right))$   

         \State $m_i \leftarrow \beta_1 m_{i-1} + (1-\beta_1) g_i$  

         \State $v_i \leftarrow \beta_2 v_{i-1} + (1-\beta_2) g_{i}^{2}$  

         \State $\hat{m}_i \leftarrow \frac{m_i}{1-\beta_1^i}$  \& $\hat{v}_i \leftarrow \frac{v_i}{1-\beta_{2}^{i}}$  


         \State $\boldsymbol{\alpha}_{i} \leftarrow \boldsymbol{\alpha}_{i-1} - \nu \frac{\hat{m}_i}{\sqrt{\hat{v}_i} + \epsilon}$

        \EndFor
       \State \textbf{return} $\boldsymbol{\alpha}$
	\end{algorithmic}
\end{algorithm}



\section{Numerical Results}\label{NR}
This section  presents numerical results to assess the effectiveness of the proposed scheme for  uni-polarized RIS beamforming. In the simulations, the 3GPP Urban Micro (UMi) model \cite{Urban} is employed to calculate all large-scale path loss values. Similar to \cite{saifglobecom_E}, we use $\sqrt{\frac{\beta_0}{(d_{u,m}^\text{U})^2}}$ and $\sqrt{\frac{\beta_0}{(d_{m,r}^\text{U})^2}}$ for $\text{UE}_u \rightarrow \text{RIS}_m$ and $\text{RIS}_m \rightarrow \text{UE}_{r}$ links, respectively, where $\beta_0$ denotes the path loss at the reference distance $d_\text{ref}=1$ m  and $d$ is the corresponding distance. Furthermore, the radiation pattern $G_{0}(\theta)$ of each RIS element is provided in \cite{Uni}. Unless stated otherwise, we use the
 parameters as listed in Table \ref{table_1}.
 
\begin{table}[t!] 
	\renewcommand{\arraystretch}{0.9}
	\caption{Simulation Parameters}
	\label{table_1}
	\centering
	\begin{tabular}{||p{1.5cm}| p{1.8cm}|p{1.5cm}| p{1.8cm}||}
		\hline
		
		\textbf{Parameter} & \textbf{Value} & \textbf{Parameter} & \textbf{Value}\\
         \hline 
         \hline
		   
         $\gamma_0^\text{U}, \gamma^\text{RIS}_{0}$ & $83, 30$ dB  & $c$ & $3\times 10^8$ m/s  \\
         \hline
		  $p, f_c$ & $1$ w, $3$ GHz & $\beta_1, \beta_2$ & $0.9, 0.999$   \\
       
       \hline 
       $\epsilon, \nu$  & $10^{-8}, 0.001$ & $\beta_0, C$ & $10^{-6}, 100$ \\

       \hline
       $B$ & $250$ KHz   & $\sigma^2_{\zeta_r}, N_0$ & $-130$ dBm \\
      \hline
      
	\end{tabular}
	
\end{table}
\subsection{GA and Validation of Analytical Expressions}
\begin{figure}[t!]  
\begin{center}
\includegraphics[width=0.85 \linewidth, draft=false]{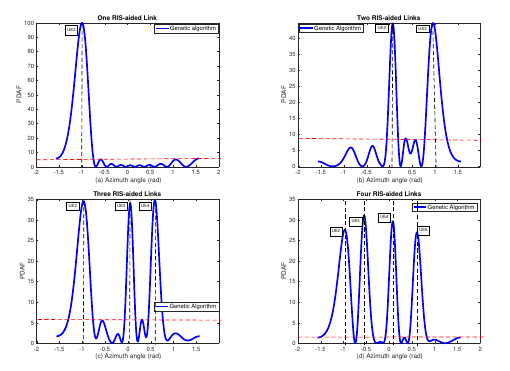}
\caption{The PDAF of the uni-polarized RIS narrowband beamforming designed using the proposed GA for (a) $\text{UE}_1 \xrightarrow{\text{RIS}} \text{UE}_{2}$, (b) $\text{UE}_1 \xrightarrow{\text{RIS}} \text{UE}_{2,3}$, (c) $\text{UE}_1 \xrightarrow{\text{RIS}} \text{UE}_{2, 3, 4}$, and (d) $\text{UE}_1 \xrightarrow{\text{RIS}} \text{UE}_{2,3,4,5}$.}
   \label{fig2}
\end{center}
\end{figure}

First, we assess the performance of the proposed GA for designing multiple narrowband beamforming per RIS towards azimuth angles of interest. Fig.~\ref{fig2} plots the PDAF versus the azimuth angle for four scenarios of RIS-aided links per the RIS, i.e.,  $U_m=1, 2, 3, 4$. For plotting Fig.~\ref{fig2}, we consider a given network of $1$ transmitting UE (UE$_1$), $1$ RIS, and $4$ receiving UEs (UE$_2$, UE$_3$, UE$_4$, UE$_5$) with fixed 2D Cartesian coordinates. Fig.~\ref{fig2} shows that the proposed GA effectively generates multiple narrowbeam RIS-aided links  towards the desired azimuth angles of the receiving UEs with the maximum PDAF. Meanwhile, it minimizes the PDAF of the other range of the azimuth angle, making the interference term in \eref{SINR} very small.

\begin{figure}[t!]  
\begin{center}
\includegraphics[width=0.85 \linewidth, draft=false]{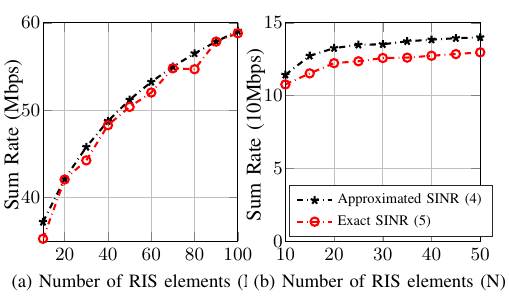}
\caption{Approximated and exact sum rates for (a) $U_m=1$ with  $\text{UE}_1 \xrightarrow{\text{RIS}_1} \text{UE}_{3}$ and $\text{UE}_2 \xrightarrow{\text{RIS}_2} \text{UE}_{4}$ and (b) $U_m=2$ with $\text{UE}_1 \xrightarrow{\text{RIS}_1} \text{UE}_{3, 6}$ and $\text{UE}_2 \xrightarrow{\text{RIS}_2} \text{UE}_{4, 5}$.}
   \label{fig3}
\end{center}
\end{figure}

In Fig.~\ref{fig3}, we plot the sum rate     to validate the tightness of the exact SINR and the SINR formulated by the approximation, as given in \eref{SINR} and \eref{app}, respectively.  For plotting Fig.~\ref{fig3}, we consider  $2$ transmitting UEs (UE$_1$, UE$_2$), $2$ RISs, and $4$ receiving UEs with fixed 2D Cartesian coordinates. From the figure, we note that the approximated sum rates for the UEs slightly match the exact sum rates for all values of $N$ for $U_m=1$, while the approximate sum rate closely aligns with the exact sum rate for  $U_m=2$. Specifically, for $U_m=2$, there is a slight reduction in  the exact sum rate since the non-aligned PDAFs are not completely suppressed as shown in Fig.~\ref{fig2}(b).
Overall, this highlights the superiority of the proposed  GA for designing unique phase shift representation that generates narrowband beamforming.  These results justify our assumption to ignore the impact of non-aligned PDAFs.

\subsection{Network Connectivity}
This subsection compares the proposed scheme  with the Perturbation (near-optimal) \cite{saifglobecom_E} and the semidefinite programming (SDP) \cite{8292633, 4786516, 4657335}, where each scheme has one narrowbeam link from each RIS. To further study the performance, we consider a network  of $2M$ small distributed RISs,  each RIS has $N/2$ elements to create a single narrowbeam link. 

\begin{figure}[t!]  
\begin{center}
\includegraphics[width=0.75 \linewidth, draft=false]{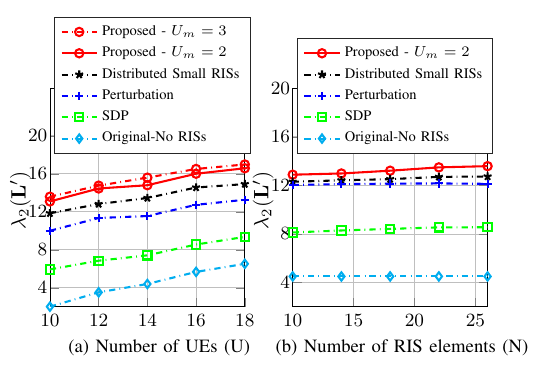}
\caption{Average network connectivity versus  (a) number of UEs $U$ and (b) number of RIS elements $N$.}
   \label{fig4}
\end{center}
\end{figure}

Fig.~\ref{fig4} shows the network connectivity versus (a) \ac{the} number of UEs $U$ for $N=10$ and (b) \ac{the} number of RIS elements $N$ for $U=10$. From Fig.~\ref{fig4}, we observe that the proposed scheme significantly outperforms the Perturbation, the SDP, and the original schemes. Additionally, the proposed scheme outperforms the scheme with distributed small RISs consisting of $6$ RISs, each having $N/2$ elements. This demonstrates that utilizing the entire RIS to generate multiple narrowband beamforming using the proposed GA is better than physically dividing the RIS into smaller ones, each generating weak signal. Such improvement is not only in terms of  network connectivity but also RIS deployment cost, as deploying fewer RISs is cheaper and easier than deploying many small RISs.  Fig.~\ref{fig4}(a)  also shows that generating many RIS-aided links per RIS decreases their PDAF as observed from  Figs.~\ref{fig2}(c), \ref{fig2}(d). This is evident when $U_m=3$; the network connectivity improvement does not increase significantly compared to $U_m=2$, indicating that the generated links are weak and the network becomes saturated in terms of link addition.

\section{Conclusion}\label{C}
This paper proposes a novel GA-based method that enables RIS to generate multiple narrowband beamforming towards desired azimuth angles of UEs through a unique phase shift determination. The proposed method shows that enabling the RIS to establish multiple narrowbeam links  rather than a single narrowbeam link markedly enhances network connectivity. The findings also show the effectiveness of the proposed method in achieving robust and connected D2D networks, demonstrating their potential  compared to the scenario where many smaller RISs are deployed, each has one RIS-aided link.

\end{document}